\documentclass[superscriptaddress,twocolumn,aps,prl]{revtex4}
\begin{document}
\title{Comment on "Can one predict DNA Transcription Start Sites by Studying Bubbles?"}
\author{C.H. Choi} \author{A. Usheva }
\affiliation{Beth Israel Deaconess Medical Center and Harvard Medical School,
99 Brookline Avenue, Boston, Massachusetts 02215}
\author{G. Kalosakas}  
\affiliation{Max Planck Institute for the Physics of Complex Systems, N\"othnitzer Str. 38, D-01187 Dresden, Germany}
\author{K. \O. Rasmussen} \author{A. R. Bishop}
\affiliation{Theoretical Division and Center for Nonlinear Studies,
Los Alamos National Laboratory, Los Alamos, New Mexico 87545}
\maketitle

Recently, van Erp {\it et al.} published an article \cite{svin} presenting conclusions which contradict earlier
works by us \cite{donald,george}. We believe this criticism to be 
misguided - indeed the authors conclusions largely substantiate the original 
discovery. This needs to be clarified so that the broader community is not mislead.

In our earlier work we provided experimental 
and theoretical evidence that functionally important sites for transcription can coincide with 
thermally induced openings of double-stranded DNA. The comment by van Erp {\em et al.} regarding 
the veracity of our experiments is without foundation. The theoretical basis for these studies was provided 
by the simple Peyrard-Bishop-Dauxois (PBD) model. Using parameters already established in the literature, we performed
Langevin simulations on the PBD model for a few specific viral DNA sequences. The object of our simulations was to 
establish whether there were certain regions in these highly heterogeneous real sequences that were 
more prone to sustain large thermally induced openings ('bubbles') of the double stranded 
molecule. Our simulations indicated that there indeed were such regions, and we experimentally verified 
their existence using 
the S1 nuclease technique.  Based on these combined theoretical and experimental findings 
we made the {\em observation} that in these viral sequences the 
regions sustaining the large bubbles coincided with known binding sites active in 
transcription events. These included, but were not limited 
to the transcription initiation site itself. Based on these observations, we concluded that it might 
be possible more generally to identify DNA functionally active sites, including transcription initiation sites, by studying 
the thermal fluctuations (particularly large amplitude coherent openings) of the double strand. It is very important 
to note that in fact these observations and speculations 
could have been made entirely based on experimental evidence.  We were, however, fortunate to also possess a model (PBD)
which sufficiently contains essential entropic ingredients to accurately guide the location of 
openings.

The claim of van Erp {\em et al.} is that our simulation technique was inadequate and therefore the entire work is 
flawed. As in our own independent elaborations \cite{zoi} of our original discovery, van Erp {\it et al.} assume that 
thermodynamic equilibrium averages are sufficient and perform explicit integrations 
of the integrals involved in the partition function for the above model. 
It is therefore correct that their and our \cite{zoi}
results with respect to obtaining thermodynamic 
averages are more accurate than our 
initial results.  However, the two crucial aspects of our initial findings 
are confirmed by these studies of thermodynamical 
equilibrium properties:
\vspace{-0.2cm}
\begin{enumerate}
\item In all three of the viral sequences examined by van Erp {\em et al.} they find the regions which sustain large 
bubbles to be precisely those we originally identified \cite{donald,george}, and these regions indeed include 
the sites active during transcription.
\vspace{-0.2cm}
\item van Erp {\em et al.} find that the two base pair mutation of the AAVP5 promoter causes a significant 
suppression of large thermal fluctuations at the former transcription sites, exactly as we reported earlier \cite{donald,george} to be the case. 
\end{enumerate}
\vspace{-0.2cm}
We emphasize again that these observations were strongly supported by experiments (notably 
missing in the work of van Erp {\it et al.}). It is indeed true that our subsequent 
studies \cite{zoi} have found that the relevant quantities for function is likely to be 
the probability of bubbles of {\it specific} sizes -- presumably associated with physical dimensions
of, e.g., transcription machinery -- but this does not dilute our primary discovery.  

Van Erp {\it et al.} make one valid point in regard to our earlier work:
We published results on a "control" sequence, which consisted of a 
non-promoter containing a similar number of base pairs as the promoter
sequences. The results we showed for this case
were unfortunate as they indicated
that no bubbles occurred in this sequence. More accurate
Langevin results do show the occurrence of bubbles in this sequence, as noted by van Erp {\em et al.} This may indicate that a human coding gene
was a poor choice for the control sequence, and that our scenario 
is best suited to viral sequences or, again that the specific bubble sizes are key.
In any case this does not affect the validity of our original results
for the active promoters.

In conclusion the results of van Erp {\em et al.} in essence confirm our initial discovery: the PBD model is 
indeed an accurate guide to the location of experimentally-identified active openings. Clearly this simple model
must be further augmented to describe either fully realistic dynamics, or how biological machinery (such as RNA polymerase)
engages these active regions. We are working to implement these augmentations.

\end{document}